\documentclass[11pt, a4paper]{article}
\usepackage{amssymb}
\usepackage[english]{babel}
\usepackage{amsmath,amsthm,amssymb,amsfonts}
\usepackage[colorlinks = true]{hyperref}
\usepackage{makeidx}
\usepackage[pdftex]{graphicx, xcolor}
\usepackage[all]{xy}
\usepackage{extarrows}
\usepackage[new]{old-arrows}
\usepackage{cancel}
\usepackage[top=3.80cm, bottom=2cm, left=2cm, right=2cm]{geometry}
\usepackage[utf8]{inputenc}
\usepackage[T1]{fontenc}
\usepackage{tikz}
\usetikzlibrary{shapes.geometric, arrows.meta, calc, positioning}

\usepackage{needspace}
\allowdisplaybreaks

\theoremstyle{plain} 
\newtheorem{thm}{Theorem}[section]

\newtheorem*{theorem*}{Theorem} 
\newtheorem*{prop*}{Proposition}

\theoremstyle{definition} 
\newtheorem{defn}[thm]{Definition}

\theoremstyle{definition} 
\newtheorem{oss}[thm]{Remark} 
\newtheorem{ex}[thm]{Example}

\DeclareMathSymbol{\Eta}{\mathalpha}{operators}{"48}

\title{\Large{\bfseries Towards General Relativity as a generalized Yang-Mills theory}}
\author{\normalsize Hartwig Winterroth\thanks{Corresponding author.}
\\
{\footnotesize Department of Mathematics, University of Torino}
\\
{\footnotesize via Carlo Alberto 10, 10123 Torino, Italy}
\\
{\footnotesize e--mail: {\sc hartwig.winterroth@unito.it}}
\\
\\
\normalsize Lorenzo Fatibene
\\ 
{\footnotesize Department of Mathematics, University of Torino}
\\
{\footnotesize via Carlo Alberto 10, 10123 Torino, Italy}
\\
{\footnotesize Istituto Nazionale di Fisica Nucleare (INFN), Sezione di Torino}
\\
{\footnotesize via Pietro Giuria 1, 10125 Torino, Italy}
\\
{\footnotesize e--mail: {\sc lorenzo.fatibene@unito.it}}
}
\date{}
\begin{document}

\maketitle

\begin{abstract}

\noindent As an application of the generalized principal bundle theory to covariant Lagrangian field theories, we aim at the development of an instance of {\em generalized gauge theories}, with the prospect of a unifying language for Yang-Mills theories and General Relativity. After reviewing the basic definitions of Lie group fiber bundles and generalized principal bundles, we provide horizontal lift and local characterizations of Lie group fiber bundle connections and generalized principal connections. Subsequently, we consider in the framework of classical field theories the kinematics and dynamics sides of generalized principal connections, which lead to a proposed notion of {\em generalized Yang-Mills theories}. Finally, we show how vector bundles are examples of generalized principal bundles and that a generalized principal connection on a vector bundle is an affine connection given in terms of basic soldering forms. We are able to recover, under appropriate assumptions, the {\em Vielbein} formulation of (vacuum) General Relativity in this setting, hinting at a (generalized) gauge theory of gravity.
\end{abstract}


\medskip


\section{Introduction}\label{sec 1}

At first sight, this contribution is concerned with a simple question: in which sense (vacuum) General Relativity can be seen as a gauge theory? Nonetheless, this specific problem actually has a rich and long history which led to many different perspectives, approaches and answers, and has also a crucial relevance for the (still unaccomplished) quantization of gravity.

Already in \cite{UTIYAMA (1956)} some analogies between Yang-Mills fields (principal connections) and linear connections (or more exactly Levi-Civita connections) concerning their transformation laws and the behaviour of the respective curvatures were pointed out. At the time of \cite{TRAUTMAN (1980)} (which also includes an excellent summary of the state of the art, until 1980, on the topic) it was clear that the question itself is ill-posed: the answer strongly depends on what we {\em precisely} mean by gauge theory and on which equivalent formulation of General Relativity we are dealing with (metric or frame formalism, Levi-Civita or independent connection, teleparallel equivalent and so on).

On the other hand, it had already become clear that principal bundles and variational calculus on fiber bundles (i.e.\,\,{\em classical field theories}) constitute the right mathematical framework for gauge theories (without quantizations), an observation that resulted in positive answers to our question both in \cite{FERRARIS (1982)} and in \cite{TRAUTMAN (1980)}.

However, these conclusions did not close the issue. In fact, addressing the topic gave inspiration for many new developments in the literature, both in mathematics and in physics. As some, not exhaustive, examples we can cite the Holst model (a modification of the frame-affine formulation of General Relativity used in the Loop Quantum Gravity approach) and also the recent papers \cite{BREZINA (2026)}, on teleparallel gravity, and \cite{HO (2016)}, on the interpretation of a class of gravitational theories as gauge theories.

\medskip

We will consider the problem from a different angle: instead of using the principal bundle setting and trying to somehow reconcile it with gravitation (the most common approach in many of the aforementioned references and what we can call the {\em bottom-up} approach), we want to consider a more general scheme of which both usual principal bundle theory and General Relativity can be seen as straightforward specializations (what we can call the {\em top-down} approach).

The question then becomes: how far should we go in the enlargement of the fundamental geometric structure (i.e.\,\,principal bundles)? A promising choice is to consider the notion of {\em generalized principal bundle}, and the associated one of {\em generalized principal connection}, as proposed by Castrillón López and Rodríguez Abella in \cite{CASTRILLON LOPEZ (2023)} (and by Fischer in \cite{FISCHER (2022)} with an independent formulation) and which we addressed also in \cite{FATIBENE (2026A)}. 

As a side note, we underline here that generalized principal bundles have a Lie groupoid-theoretic interpretation as it turns out that they are particular examples of {\em principal groupoid bundles}, see \cite{MOERDIJK (2003)} for more details on the latter concept.

\medskip

The intended aim of our approach is to develop an instance of {\em generalized gauge theories}, anticipating a unifying language for different kinds of covariant field theories, with emphasis on Yang-Mills theories and General Relativity. Specifically, we will put forward in Section \ref{sec 5} the concept of {\em generalized Yang-Mills theories}, which might perform this task.

\medskip

A more detailed and extended version of this contribution, including full mathematical results and proofs, is currently in preparation, see \cite{FATIBENE (2026B)}.

\medskip

In the following, we denote by $(P, M, \pi_{P, M}, F)$ a smooth fiber bundle such that $P$ is the total space, $M$ is the base space, $F$ is the standard fiber and $\pi_{P, M} \colon P \rightarrow M$ is the projection. Moreover, the notation $(x^\mu, y^i)$ will stand for a system of fibered coordinates on $P$, where $\mu \in\{1, \dots, \mathrm{dim}(M)\}$ and $i \in\{1, \dots, \mathrm{dim}(F)\}$. Given two manifolds $M$ and $N$, we denote the tangent map of a smooth map $f \colon M \rightarrow N$ at a point $x \in M$ by $T_x(f) \colon T_{x}(M) \rightarrow T_{f(x)}(N)$. The Einstein summation convention for repeated indices regarding local coordinates is assumed. 

\section{Lie group fiber bundles and generalized principal bundles}\label{sec 2}

This section is dedicated to a review of the geometric objects that are used in our {\em top-down} approach to the topic: generalized principal bundles. As mentioned in the Introduction, one should refer mainly to \cite{CASTRILLON LOPEZ (2023)} (and also to \cite{CASTRILLON LOPEZ (2024)}). The results presented here share the same notation that we have adopted in \cite{FATIBENE (2026A)}.

\medskip

Firstly, consider the following definition (upon which the definition of generalized principal bundle depends):
\begin{defn}\label{defn 2.1}
A {\em Lie group fiber bundle with typical fiber} $G$ is a fiber bundle $(\mathcal{G}, M, \pi_{\mathcal{G}, M}, G)$, where $G$ is a Lie group, such that:
\begin{enumerate}
\item $\forall \, x \in M$, the fiber $\mathcal{G}_x=(\pi_{\mathcal{G}, M})^{-1}(x)$ is equipped with a Lie group structure.
\item $\forall \, x \in M$, there is a neighbourhood $U \subseteq M$ of $x$ and a local trivialization:
\begin{align*}
\psi \colon (\pi_{\mathcal{G}, M})^{-1}(U) \longrightarrow U \times G
\end{align*}
preserving the Lie group structure fiberwise, i.e.\,\,$\psi|_{\mathcal{G}_x} \colon \mathcal{G}_x \longrightarrow \{x\} \times G \equiv G$ is a Lie group isomorphism.
\end{enumerate}
\end{defn}
The main consequences of these axioms are that the transition functions of $(\mathcal{G}, M, \pi_{\mathcal{G}, M}, G)$ take values in $\mathrm{Aut}(G)$ and that there exist fibered coordinates $(x^\mu, g^I)$ on $(\mathcal{G}, M, \pi_{\mathcal{G}, M}, G)$ with transformation laws:
\begin{align}\label{1}	
		\left\{
		\begin{array}{l}
			x'^\mu=x'^\mu(x)
			\\
			g'^I=G^I(x, g)
		\end{array}
		\right.	
\end{align}
satisfying unitarity, $G^I(x, e)=e^I$, and multiplicativity, $G^I\Big(x, \pi(g, h)\Big)=\pi^I\Big(G(x, g), G(x, h)\Big)$, conditions. Here $e$ indicates the identity element and $\pi \colon G \times G \rightarrow G$ is the multiplication in the Lie group $G$.

\begin{oss}\label{oss 2.2}
One can observe that a Lie group fiber bundle is in general not a principal bundle, whereas vector bundles are examples of these bundles with typical fiber some $(\mathbb{R}^l, +)$.
Moreover, given a principal bundle $(P, M, \pi_{P, M}, G)$, its adjoint bundle $(Ad(P), M, \pi_{Ad(P), M}, G)$ (of which the sections are in a one to one correspondence to the group of gauge transformations of the principal bundle) is an instance of Lie group fiber bundle, although not all Lie group fiber bundles are adjoint bundles, as already noted by Mackenzie in \cite{MACKENZIE (1989)}.
\end{oss}

By transferring the notion of action of a Lie group on a smooth manifold to the present context, it is possible to consider the concept of {\em fibered action}:
\begin{defn}\label{defn 2.3}
A {\em (right) fibered action} of a Lie group fiber bundle $(\mathcal{G}, M, \pi_{\mathcal{G}, M}, G)$ on a fiber bundle $(P, M, \pi_{P, M}, F)$ is a smooth bundle morphism $\Phi \colon P \times_M \mathcal{G} \longrightarrow P$ over the identity ${id}_M \colon M \longrightarrow M$, where $P \times_M \mathcal{G}$ is the fibered product of the two bundles, such that:
\begin{enumerate}
\item $\forall \, p \in P$, $\Phi(p, e_x)=p$, with $\pi_{P, M}(p)=x$, where $e_x$ is the identity element of the Lie group $\mathcal{G}_x$.
\item $\forall \, (p, \gamma_1), (p, \gamma_2) \in P \times_M \mathcal{G}$, $\Phi\Big(p, \mathcal{M}(\gamma_1, \gamma_2)\Big)=\Phi\Big(\Phi(p, \gamma_1), \gamma_2\Big)$, where the {\em multiplication map} $\mathcal{M} \colon \mathcal{G} \times_M \mathcal{G} \longrightarrow \mathcal{G}$ is defined by taking the product of elements in the fibers of $\mathcal{G}$.
\end{enumerate}
A fibered action is {\em free} if $\Phi(p, \gamma)=p$, for some $(p, \gamma) \in P \times_M \mathcal{G}$, then $\gamma=e_x$, with $\pi_{P, M}(p)=x$, and is {\em proper} if the smooth map $\Theta \colon P \times_M \mathcal{G} \longrightarrow P \times_M P \colon (p, \gamma) \longmapsto \Big(p, \Phi(p, \gamma)\Big)$ is a proper map.
\end{defn}

Given a fibered action $\Phi$ of $(\mathcal{G}, M, \pi_{\mathcal{G}, M}, G)$ on $(P, M, \pi_{P, M}, F)$, it is possible to introduce an equivalence relation $\sim_{\mathcal{G}}$ on $P$ declaring:
\begin{align*}
p_1 \sim_{\mathcal{G}} p_2  \, \Longleftrightarrow \, \exists \, \gamma \in \mathcal{G} \, | \, \Phi(p_1, \gamma)=p_2
\end{align*}
for $p_1, p_2 \in P$. One can then define the quotient space $\mathcal{S}=P/\mathcal{G}=P/{\sim_{\mathcal{G}}}$ for which it holds that:
\begin{thm}\label{thm 2.4}
Given a free and proper fibered action $\Phi$ of a Lie group fiber bundle $(\mathcal{G}, M, \pi_{\mathcal{G}, M}, G)$ on a fiber bundle $(P, M, \pi_{P, M}, F)$, then $\mathcal{S}$ admits a unique smooth structure such that $(P, \mathcal{S}, \pi_{P, \mathcal{S}}, G)$ is a fiber bundle.
\end{thm}

It is important to stress that this theorem essentially amounts to a generalization of the well-known Quotient Manifold Theorem. 

\needspace{3\baselineskip}
At this point it is possible to define generalized principal bundles:
\begin{defn}\label{defn 2.5}
A {\em generalized principal bundle with structure Lie group fiber bundle} $\mathcal{G}$ is a fiber bundle $(P, M, \pi_{P, M}, F)$ equipped with a free and proper right fibered action $\Phi \colon P \times_M \mathcal{G} \longrightarrow P$ of a Lie group fiber bundle $(\mathcal{G}, M, \pi_{\mathcal{G}, M}, G)$. We also say that the generalized principal bundle is the fiber bundle $(P, \mathcal{S}, \pi_{P, \mathcal{S}}, G)$ obtained from $(P, M, \pi_{P, M}, F)$ through Theorem \ref{thm 2.4} (as stated in \cite{CASTRILLON LOPEZ (2023)}).
\end{defn}

Note how principal bundles are constructed following the same path of generalized principal bundles, but we start with a smooth manifold $P$ instead of a fiber bundle $(P, M, \pi_{P, M}, F)$ and we consider just a Lie group $G$ instead of a Lie group fiber bundle $(\mathcal{G}, M, \pi_{\mathcal{G}, M}, G)$, then we take a free and proper right action $m \colon P \times G  \rightarrow P$ instead of a free and proper right fibered action $\Phi \colon P \times_M \mathcal{G} \rightarrow P$. Thanks to the usual Quotient Manifold Theorem, $(P, S, \pi_{P, S}, G)$ is a fiber bundle, the {\em principal bundle}, where $S=P/G$ is the quotient manifold defined by $m$.

\begin{ex}\label{ex 2.6}
It turns out that principal bundles and Lie group fiber bundles can be seen as examples of generalized principal bundles. Since we have noted in Remark \ref{oss 2.2} that Lie group fiber bundles are not principal bundles, we deduce that the notion of generalized principal bundle eventually extends the one of principal bundle.

Precisely, considering the fibered action $\Phi_{m} \colon P \times_M (M \times G) \rightarrow P \colon \Big(p, (x, g)\Big) \mapsto m(p, g)$ of the trivial Lie group fiber bundle $(M \times G, M, \pi_{M \times G, M}, G)$ on a principal bundle $(P, M, \pi_{P, M}, G)$ with right action $m$, it holds that the corresponding generalized principal bundle is the principal bundle $(P, M, \pi_{P, M}, G)$ itself. Considering instead the fibered action of a Lie group fiber bundle $(\mathcal{G}, M, \pi_{\mathcal{G}, M}, G)$ on itself given by the multiplication map $\mathcal{M} \colon \mathcal{G} \times_M \mathcal{G} \rightarrow \mathcal{G}$, it holds that the corresponding generalized principal bundle is the Lie group fiber bundle $(\mathcal{G}, M, \pi_{\mathcal{G}, M}, G)$ itself.
\end{ex}

In \cite{FATIBENE (2026A)} we used Theorem \ref{thm 2.4} to establish on a generalized principal bundle $(P, \mathcal{S}, \pi_{P, \mathcal{S}}, G)$ {\em generalized principal bundle coordinates} $(x^\mu, \sigma^\Theta, g^I)$, i.e.\,\,fibered coordinates adapted to the generalized principal bundle structure where $(x^\mu, \sigma^\Theta)$ are fibered coordinates on $\mathcal{S}$, which have transformation laws:
\begin{align}\label{2}	
                 \left\{
		\begin{array}{l}
			x'^\mu=x'^\mu(x)
			\\
			\sigma'^\Theta=\Sigma^\Theta(x, \sigma)
			\\
			g'^I=\pi^I\Big(\varphi(x, \sigma), G(x, g)\Big)
		\end{array}
		\right.
\end{align}
where $\pi^I$ are the local expressions of the product in the Lie group $G$, $\varphi(x, \sigma)=\varphi^A(x, \sigma)$ are the local expressions of a function $\varphi$ (characterizing the generalized principal bundle) and $G(x, g)=G^B(x, g)$, as well as $x'^\mu(x)$, are the same local expressions as in \eqref{1}, with $\mu \in \{1, \dots, m\}$, $\Theta \in \{1, \dots, n-l\}$, $A, B, I \in \{1, \dots, l\}$, $\mathrm{dim}(M)=m$, $\mathrm{dim}(F)=n$, $\mathrm{dim}(G)=l$.

\section{Lie group fiber bundle connections and generalized principal connections}\label{sec 3}

In \cite{FATIBENE (2026A)} we have also proposed a straightforward way to characterize {\em Lie group fiber bundle connections}, as well as {\em generalized principal connections}, in terms of horizontal lifts (see for example \cite{FATIBENE (2003)} and \cite{KOBAYASHI (1963)} for more details) and hence we present here some equivalent definitions (to the ones appeared in \cite{CASTRILLON LOPEZ (2023)}) of these notions based on such characterizations.

\medskip

Lie group fiber bundle connections are needed in order to define generalized principal connections and are constructed respecting the inherent multiplicative structure on a Lie group fiber bundle:
\begin{defn}\label{defn 3.1}
A {\em Lie group fiber bundle connection} on the Lie group fiber bundle $(\mathcal{G}, M, \pi_{\mathcal{G}, M}, G)$ is a connection $\eta$, given by the collection of the horizontal lifts $\eta_{\gamma} \colon T_{x}(M) \longrightarrow T_{\gamma}(\mathcal{G})$, $\forall \, x \in M$, $\forall \, \gamma \in \mathcal{G}$, 

\needspace{3\baselineskip}
\noindent with $\pi_{\mathcal{G}, M}(\gamma)=x$, such that:
\begin{enumerate}
\item $\forall \, x \in M$, $\eta_{e_x}=T_{x}(1)$, where $1 \colon M \rightarrow \mathcal{G}$, $x \mapsto e_x \in \mathcal{G}_x$ is the {\em unit section} (recall that $e_x$ is the identity element of the Lie group $\mathcal{G}_x=(\pi_{\mathcal{G}, M})^{-1}(x)$).
\item $\forall \, (\gamma, \delta) \in \mathcal{G} \times_M \mathcal{G}$, $\eta_{\gamma \cdot \delta}=T_{(\gamma, \delta)}(\mathcal{M})\Big(\eta_{\gamma}(\cdot), \eta_{\delta}(\cdot)\Big)$, where $\mathcal{M}$ is the multiplication map.
\end{enumerate}
\end{defn}
It is subsequently possible to use Definition \ref{defn 3.1} to show that in fibered coordinates $(x^\mu, g^I)$ satisfying \eqref{1} a Lie group fiber bundle connection can be locally written as:
\begin{align*}
\eta=dx^\mu \otimes \Big(\partial_{\mu}-\eta^I_\mu(x, g)\partial_{I}\Big)
\end{align*}
where we set $\partial_{\mu}=\frac{\partial}{\partial x^\mu}$ and $\partial_{I}=\frac{\partial}{\partial g^I}$, with the connection coefficients $\eta^I_\mu(x, g)$ satisfying the conditions:
\begin{enumerate}
\item $\eta^I_\mu(x, e)=0$, where $e$ is the identity element of the Lie group $G$.
\item $\forall \, g, h \in G$, $\eta^I_\mu\Big(x, \pi(g, h)\Big)=\eta^A_\mu(x, g)\partial^1_A\pi^I(g, h)+\eta^A_\mu(x, h)\partial^2_A\pi^I(g, h)$, where $\pi^I(g, h)$ are the local expressions of the product in the Lie group $G$ and $\partial^1_A\pi^I(g, h)$, $\partial^2_A\pi^I(g, h)$ are the partial derivatives regarding $g^A$ and $h^A$, respectively.
\end{enumerate}

We can then use the same type of horizontal lift characterization to define generalized principal connections:
\begin{defn}\label{defn 3.2}
Let $\eta$ be a connection on the Lie group fiber bundle $(\mathcal{G}, M, \pi_{\mathcal{G}, M}, G)$. A {\em generalized principal connection} on the generalized principal bundle $(P, \mathcal{S}, \pi_{P, \mathcal{S}}, G)$ associated to $\eta$ is a connection $\tau$, given by the collection of the horizontal lifts $\tau_{p} \colon T_{[p]_{\mathcal{G}}}(\mathcal{S}) \longrightarrow T_{p}(P)$, $\forall \, [p]_{\mathcal{G}} \in \mathcal{S}$, $\forall \, p \in P$, with $\pi_{P, \mathcal{S}}(p)=[p]_{\mathcal{G}}$, such that:
\begin{align*}
\tau_{\Phi(p, \gamma)}=T_{(p, \gamma)}(\Phi)\Big(\tau_{p}(\cdot), \eta_{\gamma}\Big(T_{[p]_{\mathcal{G}}}(\pi_{\mathcal{S}, M})(\cdot)\Big)\Big), \, \forall \, (p, \gamma) \in P \times_M \mathcal{G}
\end{align*}
where $\Phi$ is the relevant right fibered action.
\end{defn}

\begin{oss}\label{oss 3.3}
It holds that the connection $\eta$ on the Lie group fiber bundle $(\mathcal{G}, M, \pi_{\mathcal{G}, M}, G)$ fixed in Definition \ref{defn 3.2} is actually forced to be a Lie group fiber bundle connection in the sense of Definition \ref{defn 3.1}.
\end{oss}

Again, it is possible to show, using Definition \ref{defn 3.2}, that in generalized principal bundle coordinates $(x^\mu, \sigma^\Theta, g^I)$ satisfying \eqref{2} a generalized principal connection can be locally written as:
\begin{align*}
\tau=dx^\mu \otimes \Big(\partial_{\mu}-A^I_\mu(x, \sigma, g)\partial_{I}\Big)+d\sigma^\Theta \otimes \Big(\partial_{\Theta}-A^I_\Theta(x, \sigma, g)\partial_{I}\Big)
\end{align*}
where we set $\partial_{\mu}=\frac{\partial}{\partial x^\mu}$, $\partial_{\Theta}=\frac{\partial}{\partial \sigma^\Theta}$ and $\partial_{I}=\frac{\partial}{\partial g^I}$, with the connection coefficients $A^I_\mu(x, \sigma, g)$ and $A^I_\Theta(x, \sigma, g)$ satisfying the conditions:
\begin{enumerate}
\item $\forall \, g, h \in G$, $A^I_\mu\Big(x, \sigma, \pi(g, h)\Big)=A^A_\mu(x, \sigma, g)\partial^1_A\pi^I(g, h)+\eta^A_\mu(x, h)\partial^2_A\pi^I(g, h)$, where $\eta^A_\mu(x, h)$ are the connection coefficients of the fixed Lie group fiber bundle connection.
\item $\forall \, g, h \in G$, $A^I_\Theta\Big(x, \sigma, \pi(g, h)\Big)=A^A_\Theta(x, \sigma, g)\partial^1_A\pi^I(g, h)$.
\end{enumerate}

\section{Generalized principal connection coefficients and curvature}\label{sec 4}

The fundamental idea of this section is to follow and extend the principal bundle case where the principal connection coefficients on a principal bundle $(P, S, \pi_{P, S}, G)$ are seen to depend solely on the coordinates on the base space $S$ (see for example \cite{FATIBENE (2003)}). More precisely, using the results presented in Section \ref{sec 3}, it is possible to rewrite a generalized principal connection locally as:
\begin{align*}
\tau&=dx^\mu \otimes \Big(\partial_{\mu}-\Big[A^A_\mu(x, \sigma, e)\partial^1_A\pi^I(e, g)+\eta^I_\mu(x, g)\Big]\partial_{I}\Big)+d\sigma^\Theta \otimes \Big(\partial_{\Theta}-A^A_\Theta(x, \sigma, e)\partial^1_A\pi^I(e, g)\partial_{I}\Big)
\end{align*}
which suggests to define the {\em generalized principal connection coefficients} on the base space $\mathcal{S}$ in the following way:
\begin{align}
A^A_\mu(x, \sigma)&=A^B_\mu(x, \sigma, e)\bar{T}^A_B \label{3}
\\
A^A_\Theta(x, \sigma)&=A^B_\Theta(x, \sigma, e)\bar{T}^A_B \label{4}
\end{align}
once fixed a basis $\{T_A\}$ of $\mathfrak{g}$, the Lie algebra of $G$, and once we denote by $\bar{T}^A_B$ the inverse of the change of basis matrix $T^B_A$:
\begin{align*}
T_A=T^B_A\left.\partial_{B}\right|_{g=e}=T^I_A\left.\frac{\partial}{\partial g^B}\right|_{g=e}
\end{align*}
where $\frac{\partial}{\partial g^I}$ are the vector fields associated with $g^I$, intended as coordinates on $G$.

\medskip

Specializing now to the case of generalized principal connection the curvature tensor:
\begin{align*}
R^i_{\mu\nu}(x, y)=\frac{\partial \Gamma_{\nu}^i}{\partial x^{\mu}}(x, y)-\frac{\partial \Gamma_{\mu}^i}{\partial x^{\nu}}(x, y)+\Gamma_{\nu}^k(x, y)\frac{\partial \Gamma_{\mu}^i}{\partial y^k}(x, y)-\Gamma_{\mu}^k(x, y)\frac{\partial \Gamma_{\nu}^i}{\partial y^k}(x, y)
\end{align*}
existing for any connection $\Gamma^i_\mu(x, y)$ on any fiber bundle $(P, M, \pi_{P, M}, F)$, it holds that its local expressions are:
\begin{align*}
\left\{
        \begin{array}{l}
	F^I_{\mu\nu}(x, \sigma, g)=F^A_{\mu\nu}(x, \sigma)T^B_A\partial^1_B\pi^I(e, g)+(R_\eta)^I_{\mu\nu}(x, g)
	\\
	\\
	F^I_{\mu\Theta}(x, \sigma, g)=F^A_{\mu\Theta}(x, \sigma)T^B_A\partial^1_B\pi^I(e, g)
	\\
	\\
	F^I_{\Theta\mu}(x, \sigma, g)=F^A_{\Theta\mu}(x, \sigma)T^B_A\partial^1_B\pi^I(e, g)
	\\
	\\
	F^I_{\Theta\Lambda}(x, \sigma, g)=F^A_{\Theta\Lambda}(x, \sigma)T^B_A\partial^1_B\pi^I(e, g)
\end{array}
\right.
\end{align*}
where:
\begin{align}
F^A_{\mu\nu}(x, \sigma)&=\partial_{\mu}A^A_\nu(x, \sigma)-\partial_{\nu}A^A_\mu(x, \sigma)+c^A{}_{DE}A^D_\mu(x, \sigma)A^E_\nu(x, \sigma)+A^D_\nu(x, \sigma)\Eta^A_{D\mu}(x) \label{5}
\\
&\,\,\,\,\,\,-A^D_\mu(x, \sigma)\Eta^A_{D\nu}(x) \nonumber
\\
F^A_{\mu\Theta}(x, \sigma)&=\partial_{\mu}A^A_\Theta(x, \sigma)-\partial_{\Theta}A^A_\mu(x, \sigma)+c^A{}_{DE}A^D_\mu(x, \sigma)A^E_\Theta(x, \sigma)+A^D_\Theta(x, \sigma)\Eta^A_{D\mu}(x) \nonumber
\\
&=-F^A_{\Theta\mu}(x, \sigma) \nonumber
\\
F^A_{\Theta\Lambda}(x, \sigma)&=\partial_{\Theta}A^A_\Lambda(x, \sigma)-\partial_{\Lambda}A^A_\Theta(x, \sigma)+c^A{}_{DE}A^D_\Theta(x, \sigma)A^E_\Lambda(x, \sigma) \nonumber
\end{align}
which we call {\em generalized curvature coefficients} and where $\Eta^A_{C\alpha}(x)=T^D_C\partial_D\eta^E_\alpha(x, e)\bar{T}^A_E$.

In the preceding expressions there also explicitly appear the generalized principal connection coefficients introduced in \eqref{3} and \eqref{4}, the structure constants $c^A{}_{DE}$ of the Lie group $G$ for the fixed basis $\{T_A\}$ and the curvature coefficients:
\begin{align*}
(R_\eta)^J_{\mu\nu}(x, g)&=\partial_{\mu}\eta^J_\nu(x, g)-\partial_{\nu}\eta^J_\mu(x, g)+\eta^A_\nu(x, g)\partial_A\eta^J_\mu(x, g)-\eta^A_\mu(x, g)\partial_A\eta^J_\nu(x, g)
\end{align*}
for the fixed Lie group fiber bundle connection $\eta$.

At this point it would also be natural to see what form the {\em Bianchi identities}, holding for any connection $\Gamma^i_\mu(x, y)$:
\begin{align*}
B^i_{\alpha\mu\nu}+B^i_{\nu\alpha\mu}+B^i_{\mu\nu\alpha}=0 \quad\quad B^i_{\alpha\mu\nu}=\Big(\partial_{\alpha}-\Gamma^k_\alpha(x, y)\partial_{k}\Big)R^i_{\mu\nu}(x, y)+R^k_{\mu\nu}(x, y)\partial_k\Gamma_{\alpha}^i(x, y)
\end{align*}
take for generalized principal connections, as they have a fundamental role in field theories. For example, the homogeneous part of the Maxwell equations:
\begin{align*}
\partial^{}_{\alpha}F^{}_{\mu\nu}(x)+\partial^{}_{\mu}F^{}_{\nu\alpha}(x)+\partial^{}_{\nu}F^{}_{\alpha\mu}(x)=0
\end{align*}
where $F_{\mu\nu}(x)$ is the Faraday tensor, are the Bianchi identities for the four-potential of the Maxwell theory (or, in other words, for the $U(1)$-principal connection) $A_{\mu}(x)$.

\medskip

This particular problem, which leads to the formulation of the {\em generalized homogeneous field equations}, is extensively addressed in \cite{FATIBENE (2026B)}.

\section{Generalized Yang-Mills theories}\label{sec 5}

We are able to consider not only the kinematics side (under the name of the generalized homogeneous field equations mentioned in Section \ref{sec 4}), but also the dynamics side of generalized principal connections once we introduce into the picture the framework of {\em classical field theories}, i.e.\,\,of variational calculus on fiber bundles (see for example to \cite{FATIBENE (2003)}, \cite{GIACHETTA (2009)} and \cite{KOLAR (1993)}), which amounts to fixing a {\em configuration bundle} and a {\em Lagrangian} on it, considering then as field equations the relevant {\em Euler-Lagrange equations}. In our case the fields will be the generalized principal connection coefficients introduced in \eqref{3} and \eqref{4}.

\medskip

{\em Yang-Mills theories} can be seen as a special class of classical field theories, where the configuration bundle is set to be the {\em principal connection bundle} $\mathrm{Con}(P)$, associated with a principal bundle $(P, S, \pi_{P, S}, G)$. This is the fiber bundle of which the sections are in a one to one correspondence with principal connections on $(P, S, \pi_{P, S}, G)$ itself. We define $F^A_{\mu\nu}=A^A_{\nu, \mu}-A^A_{\mu, \nu}+c^A{}_{DE}A^D_\mu A^E_\nu$ from the fibered coordinates $(x^\mu, A^A_\mu, A^A_{\mu, \nu})$ on $J^1\mathrm{Con}(P)$, the first order jet prolongation of $\mathrm{Con}(P)$ (see \cite{SAUNDERS (1989)} for more details). Regarding the Lagrangian, we fix on $S$ a metric $g=g_{\mu\nu}(x)dx^\mu \otimes dx^\nu
$ and ask the structure group $G$ to be semisimple so that the $Ad$-invariant symmetric {\em Cartan-Killing bilinear form} $K$ on the Lie algebra $\mathfrak{g}$ is non-degenerate. On a basis $\{T_A\}$ it holds that:
\begin{align*}
K(T_A, T_B)=K_{AB}=c^D{}_{AE}c^E{}_{BD}
\end{align*}
We then choose the {\em Yang-Mills Lagrangian}:
\begin{align}\label{6}
{L}_{YM}=-\frac{1}{4}F^A_{\mu\nu}g^{\mu\alpha}(x)g^{\nu\beta}(x)F^B_{\alpha\beta}K_{AB}\sqrt{g(x)}ds
\end{align}
where $g(x)=\Big|\det\Big(g_{\mu\nu}(x)\Big)\Big|$ and $ds=dx^1 \wedge \dots \wedge dx^k$, where $k=\mathrm{dim}(S)$.

\medskip

By following the preceding definition path of Yang-Mills theories, but using generalized principal bundles instead of principal bundles, it is possible to build a {\em generalized principal connection bundle} and to fix on it a {\em generalized Yang-Mills Lagrangian}. This new Lagrangian has the same algebraic structure of \eqref{6}, but the $F^A_{\mu\nu}$ terms correspond to the generalized curvature coefficients \eqref{5} introduced in Section \ref{sec 4}.

These choices lead to a notion of {\em generalized Yang-Mills theory}, the instance of generalized gauge theories we were foreshadowing in the Introduction, which turns out to eventually include and generalize Yang-Mills theories; for more details on this we refer once more to the upcoming \cite{FATIBENE (2026B)}.

\section{Hints of a (generalized) gauge theory of gravity}\label{sec 6}

Through Remark \ref{oss 2.2} and Example \ref{ex 2.6} we can deduce that any vector bundle $(E, M, \pi_{E, M}, \mathbb{R}^l)$ is a generalized principal bundle arising from a free and proper fibered action of the vector bundle on itself. As an application of Section \ref{sec 3}, it is possible to prove that a generalized principal connection on $(E, M, \pi_{E, M}, \mathbb{R}^l)$ associated to a fixed Lie group fiber bundle connection on $(E, M, \pi_{E, M}, \mathbb{R}^l)$ itself is an {\em affine connection}.

Precisely, using vector bundle coordinates $(x^\mu, v^I)$ on $(E, M, \pi_{E, M}, \mathbb{R}^l)$ satisfying:
\begin{align*}
		\left\{
		\begin{array}{l}
			x'^\mu=x'^\mu(x)
			\\
			v'^I=G^I_A(x)v^A
		\end{array}
		\right.	
\end{align*}
where $G^I_A(x) \in \mathrm{GL}(l, \mathbb{R})$, which play the role in this case of both the fibered coordinates \eqref{1} and of the generalized principal connection coordinates \eqref{2}, it holds that the Lie group fiber bundle coefficients satisfy $\eta^I_\mu(x, v)=\eta^I_{A\mu}(x)v^A$ or, in other words, the Lie group fiber bundle connections on the fixed vector bundle are exactly the {\em linear connections} on it, as well as it holds that the connection coefficients of a generalized principal connection satisfy $A^I_\mu(x, v)=\sigma^I_\mu(x)+\eta^I_{A\mu}(x)v^A$ once set $\sigma^I_\mu(x)=A^I_\mu(x, 0)$. 

In this sense, we can conclude that the generalized principal connections on a vector bundle $(E, M, \pi_{E, M}, \mathbb{R}^l)$ are exactly the affine connections on it, where the respective underlying linear connections are the fixed Lie group fiber bundle connections on $(E, M, \pi_{E, M}, \mathbb{R}^l)$ itself (see \cite{FATIBENE (2026A)} for all the details).

\medskip

It is in particular possible to prove that $\sigma^I_\mu(x)$ transforms as:
\begin{align*}
\sigma'^I_\mu\Big(x'(x)\Big)&=\bar{J}^\nu_\mu\Big(x'(x)\Big)\sigma^A_\nu(x)G^I_A(x)
\end{align*}
where ${\bar{J}}^{\nu}_{\mu}(x')=\frac{\partial x^\nu}{\partial x'^\mu}(x')$, hence $\sigma^I_\mu(x)$ are the local expressions of a {\em basic soldering form} $\sigma$ on the vector bundle $(E, M, \pi_{E, M}, \mathbb{R}^l)$, i.e.\,\,a section of the vector bundle $(T^{*}(M) \otimes_{M} E, M, \pi_{T^{*}(M) \otimes_{M} E, M}, \mathbb{R}^{m \cdot l})$, where $m=\mathrm{dim}(M)$:
\begin{align*}
\sigma \colon M \longrightarrow T^{*}(M) \otimes_{M} E \colon x^\mu \longmapsto \Big(x^\mu, \sigma^I_\mu(x)\Big)
\end{align*}
which can be seen also as a vector bundle morphism:
\begin{align*}
\hat{\sigma} \colon T(M) \longrightarrow E \colon (x^\mu, \dot{x}^\mu) \longmapsto \Big(x^\mu, \sigma^I_\mu(x)\dot{x}^\mu\Big)
\end{align*}
since $\sigma'^I_\mu\Big(x'(x)\Big)\dot{x}'^\mu=\bar{J}^\nu_\mu\Big(x'(x)\Big)\sigma^A_\nu(x)G^I_A(x)J^\mu_\alpha(x)\dot{x}^\alpha=\delta^\nu_\alpha\sigma^A_\nu(x)G^I_A(x)\dot{x}^\alpha=G^I_A(x)\sigma^A_\nu(x)\dot{x}^\nu$, where ${J}^{\mu}_{\alpha}(x)=\frac{\partial x'^\mu}{\partial x^\alpha}(x)$ (see also \cite{MODUGNO (1991)}).

Moreover, in this case the generalized curvature coefficients \eqref{5} turn out to be of the form:
\begin{align*}
F^A_{\mu\nu}(x)&=(T_{\eta, \sigma})^B_{\mu\nu}(x)\bar{T}^A_B
\end{align*}
where $(T_{\eta, \sigma})^B_{\mu\nu}(x)=\partial_{\mu}\sigma^B_\nu(x)-\partial_{\nu}\sigma^B_\mu(x)+\sigma^C_\nu(x)\eta^B_{C\mu}(x)-\sigma^C_\mu(x)\eta^B_{C\nu}(x)$ are the {\em torsion coefficients} of the linear connection $\eta$ with respect to the basic soldering form $\sigma$ (compare again with \cite{MODUGNO (1991)}).

\medskip

We can now note that the construction presented in Section \ref{sec 5} can be slightly modified when the generalized principal bundle $(P, \mathcal{S}, \pi_{P, \mathcal{S}}, G)$ is set to be a vector bundle $(E, M, \pi_{E, M}, \mathbb{R}^l)$ over a Lorentzian manifold $M$ with $\mathrm{dim}(M)=4$, i.e.\,\,a manifold admitting metrics with signature $(3, 1)\equiv(-1, 1, 1, 1)$. In particular, we want to show that it is possible to recover the {\em Vielbein} formulation of (vacuum) General Relativity in this setting, under appropriate assumptions.

Suppose then that $\hat{\sigma} \colon T(M) \rightarrow E$ is a vector bundle isomorphism, which implies that $l=4$ and $\mathrm{det}\Big(\sigma^I_\mu(x)\Big)\neq0$. With this assumption the present framework becomes the {\em fake tangent bundle} or {\em internal space} environment used in the definition of the Vielbein formalism, the Palatini formalism and even the teleparallel equivalent of General Relativity (see for example \cite{BAEZ (2015)}). In this context the basic soldering form $\sigma$, or equivalently $\hat{\sigma}$, is called Vielbein or {\em coframe field}.

Since $M$ is a Lorentzian manifold and hence $(E, M, \pi_{E, M}, \mathbb{R}^4) \cong (T(M), M, \pi_{T(M), M}, \mathbb{R}^4)$ has a $SO(3, 1)$-reduction (i.e.\,\,from now on $G^I_A(x) \in SO(3, 1)$), we can define, as it is well-known, a metric $g$ with signature $(3, 1)$ on $M$ by setting locally:
\begin{align*}
g_{\mu\nu}(x)&=\sigma^A_\mu(x)\eta_{AB}\sigma^B_\nu(x)
\end{align*}
where $\eta_{AB}=\mathrm{diag}(-1, 1, 1, 1)$.

Instead of the generalized principal connection bundle for the vector bundle $(E, M, \pi_{E, M}, \mathbb{R}^4)$, which turns out to be isomorphic to $(T^{*}(M)^{(\times_{M})^4}, M, \pi_{T^{*}(M)^{(\times_{M})^4}, M}, \mathbb{R}^{16})$, this time we choose as configuration bundle the fiber bundle $(T^{*}(M) \otimes_{M} E, M, \pi_{T^{*}(M) \otimes_{M} E, M}, \mathbb{R}^{16})$ of which the sections are in a one to one correspondence with basic soldering forms, as seen above. It has fibered coordinates $(x^\mu, \sigma^I_\mu)$ satisfying:
\begin{align*}
                \left\{
		\begin{array}{l}
		        x'^\mu=x'^\mu(x)
		        \\
			\sigma'^I_\mu=\bar{J}^\nu_\mu\Big(x'(x)\Big) \sigma^A_\nu G^I_A(x)
		\end{array}
		\right.
\end{align*}
where $G^I_A(x) \in SO(3, 1)$ and we must keep in mind the chosen restriction $\mathrm{det}(\sigma^I_\mu)\neq0$. This fiber bundle is also known as {\em Vielbein bundle}. These coordinates induce $(x^\mu, \sigma^I_\mu, \sigma^I_{\mu, \nu})$ as fibered coordinates on $J^1(T^{*}(M) \otimes_{M} E)$.

\medskip

Now, we consider a modification of the generalized Yang-Mills Lagrangian that we have considered above by defining locally:
\begin{align*}
{L}_{\mathrm{Vielbein}}=\Big(-\frac{1}{4}F^A_{\mu\nu}g^{\mu\alpha}g^{\nu\beta}F^B_{\alpha\beta}\eta_{AB}+R\Big)\sqrt{g}ds
\end{align*}
where: 
\begin{enumerate}
\item $g_{\mu\nu}=g_{\mu\nu}(\sigma)=\sigma^A_\mu\eta_{AB}\sigma^B_\nu$, so that $g^{\mu\nu}=g^{\mu\nu}(\sigma)=\sigma_A^\mu\eta^{AB}\sigma_B^\nu$.
\item We have substituted the Cartan-Killing bilinear form $K_{AB}$ with $\eta_{AB}$.
\item $g=\Big|\det(g_{\mu\nu})\Big|$, so that $\sqrt{g}=|\sigma|=\Big|\mathrm{det}(\sigma^I_\mu)\Big|$.
\item $F^A_{\mu\nu}$ are still defined taking inspiration from the generalized curvature coefficients, this time in the vector bundle case:
\begin{align*}
F^A_{\mu\nu}&=F^A_{\mu\nu}(\sigma)=(T_{\eta(\sigma), \sigma})^B_{\mu\nu}(\sigma)\bar{T}^A_B=\Big(\sigma^B_{\nu, \mu}-\sigma^B_{\mu, \nu}+\sigma^C_\nu\eta^B_{C\mu}(\sigma)-\sigma^C_\mu\eta^B_{C\nu}(\sigma)\Big)\bar{T}^A_B
\end{align*}
where $\eta^B_{C\mu}(\sigma)=\sigma^\alpha_C\eta^\beta_{\alpha\mu}(\sigma)\sigma^B_\beta-\sigma^\alpha_C\sigma^B_{\alpha, \mu}$ are the well-known {\em spin connection} or {\em Vielbein connection} coefficients. $\eta^\beta_{\alpha\mu}=\eta^\beta_{\alpha\mu}(\sigma)$ are just the Levi-Civita connection coefficients written in terms of the Vielbein $\sigma$, through the metric $g$.
\item $R=R(\sigma)$ is the Ricci scalar written in terms of the Vielbein $\sigma$, through the metric $g$.
\item $ds=dx^1 \wedge dx^2 \wedge dx^3 \wedge dx^4$.
\end{enumerate}

Essentially, we have added to the generalized Yang-Mills Lagrangian the Einstein-Hilbert term, asking also the metric to be derived from the fundamental field given by the Vielbein $\sigma$ and assuming the Vielbein postulate for the connection. We can also readily observe that:
\begin{align*}
\eta^B_{C\mu}(\sigma)=\sigma^\alpha_C\eta^\beta_{\alpha\mu}(\sigma)\sigma^B_\beta-\sigma^\alpha_C\sigma^B_{\alpha, \mu} \, \Longleftrightarrow \, \sigma^B_{\alpha, \mu}+\sigma_\alpha^C\eta^B_{C\mu}(\sigma)=\eta^\beta_{\alpha\mu}(\sigma)\sigma^B_\beta
\end{align*}
which implies:
\begin{align*}
(T_{\eta(\sigma), \sigma})^B_{\mu\nu}(\sigma)&=\sigma^B_{\nu, \mu}-\sigma^B_{\mu, \nu}+\sigma^C_\nu\eta^B_{C\mu}(\sigma)-\sigma^C_\mu\eta^B_{C\nu}(\sigma)=\eta^\beta_{\nu\mu}(\sigma)\sigma^B_\beta-\eta^\beta_{\mu\nu}(\sigma)\sigma^B_\beta
\\
&=\Big(\eta^\beta_{\nu\mu}(\sigma)-\eta^\beta_{\mu\nu}(\sigma)\Big)\sigma^B_\beta=0
\end{align*}
since $\eta^\beta_{\alpha\mu}(\sigma)$ is the Levi-Civita connection, hence $F^A_{\mu\nu}=0$. In this sense, we have that:
\begin{align*}
{L}_{\mathrm{Vielbein}}=R|\sigma|ds={L}_{HE}
\end{align*}
or, in other words, we have actually written the Hilbert-Einstein Lagrangian in the Vielbein formalism, reaching our intended aim.

\section{Conclusion}\label{sec 7}

The construction that we have presented in Section \ref{sec 6} can be interpreted as a linguistic trick. Nonetheless, it highlights the flexibility of the generalized principal bundle theory and the well-known asymmetry between Yang-Mills theories and General Relativity given by the Vielbein $\sigma$, as noted for example in \cite{ZANELLI (2012)}. In particular, it is the first evidence that Yang-Mills theories and General Relativity might be both seen as theories for a generalized principal connection, as specializations of the generalized principal bundle theory in the direction of standard principal bundles and of vector bundles, respectively. 

\medskip

More exactly, the discussion in Section \ref{sec 6} and the notion of generalized Yang-Mills theory, proposed in Section \ref{sec 5}, have to be considered as the first two steps of a path (or, if you want, two patches of a bigger picture) leading to a more complete notion of generalized Yang-Mills theory, going in the direction of a unifying {\em top-down} language for Yang-Mills theories and General Relativity.

\section*{Acknowledgements} 

We would like to acknowledge the contribution of the local research project {\em Metodi Geometrici in Fisica Matematica e Applicazioni (2025)}, Department of Mathematics, University of Torino, and of INFN ({\em Iniziativa Specifica QGSKY} and {\em Iniziativa Specifica Euclid}). This paper is also supported by INdAM-GNFM.

\end{document}